\documentclass[10pt,aps,prl,twocolumn,superscriptaddress,preprintnumbers,floatfix,notitlepage]{revtex4-1}
\usepackage{hyperref}
\hypersetup{
    colorlinks=true,
    linkcolor=blue,
    filecolor=magenta,      
    urlcolor=blue,
}
\usepackage{amsmath,amsfonts,amssymb}
\usepackage{fancyhdr}
\usepackage{graphicx}
\usepackage{xspace}
\usepackage{rotating}
\usepackage[normalem]{ulem}
\usepackage{braket}
\usepackage{verbatim}
\usepackage{xcolor}
\usepackage{hyperref}
\usepackage[utf8]{inputenc}

\def\lsim{\mathrel{\raise.3ex\hbox{$<$\kern-.75em\lower1ex\hbox{$\sim$}}}}
\def\gsim{\mathrel{\raise.3ex\hbox{$>$\kern-.75em\lower1ex\hbox{$\sim$}}}}

\newcommand{\be}{\begin{equation}}
\newcommand{\ee}{\end{equation}}
\newcommand{\bea}{\begin{equation}\begin{aligned}}
\newcommand{\eea}{\end{aligned}\end{equation}}

\begin{document}

\title{Dark Matter Constraints from the Eccentric Supermassive Black Hole Binary OJ~287}

\author{Ahmad Alachkar}
\affiliation{King's College London, Strand, London, WC2R 2LS, United Kingdom}

\author{John~Ellis}
\affiliation{King's College London, Strand, London, WC2R 2LS, United Kingdom}
\affiliation{Theoretical Physics Department, CERN, Geneva, Switzerland}
\affiliation{National Institute of Chemical Physics \& Biophysics, R\"avala 10, 10143 Tallinn, Estonia}

\author{Malcolm Fairbairn}
\affiliation{King's College London, Strand, London, WC2R 2LS, United Kingdom}

\begin{abstract}
OJ 287 is a blazar thought to be a binary system containing a $\simeq 18$~billion solar mass 
primary black hole accompanied by a $\simeq 150$ million solar mass
secondary black hole in an eccentric orbit, which triggers electromagnetic flares
twice in every $\simeq 12$~year orbital period when it traverses the accretion disk of the primary.
The times of these emissions are consistent with the predictions of general relativity
calculated to the 4.5th post-Newtonian order. The orbit of the secondary black hole
samples the gravitational field at distances between ${\cal O}(10)$ and ${\cal O}(50)$
Schwarzschild radii around the primary, and hence is sensitive to the possible presence
of a dark matter spike around it. We find that the agreement of general-relativistic calculations with the
measured timings of flares from OJ~287 constrains the mass of such a spike to
$\lesssim 3$\% of the primary mass.\\
~~\\
KCL-PH-TH/2022-40,~CERN-TH-2022-115
\end{abstract}

\maketitle


OJ 287 
is an active galactic nucleus (AGN) situated near the ecliptic in the constellation of Cancer at a cosmological redshift
$z=0.3056$~\cite{sitko1985continuum,nilsson2010h}. It is categorised as a BL Lacertae (BL Lac) object with a relativistic jet aligned very close to our line of sight \cite{stein1976bl}. 
Due to its proximity to the ecliptic, OJ 287 was often unintentionally photographed in the past, 
with an optical database dating back to 1888~\cite{1988ApJ...325..628S,2001ESASP.459..295H,visvanathan1973variations,abdo2010spectral,hudec2013historical},
in addition to dedicated observing programmes~\cite{takalo1996reflections,goyal2018stochastic,hudec2013historical}.
This huge data set contains quasi-periodic pairs of electromagnetic
flares every $\sim 12$~years \cite{valtonen2006predicting}, which are explained by Lehto and Valtonen~\cite{lehto1996oj,sundelius1997numerical} as impacts of a secondary black hole (BH) with the accretion disk of the primary BH as it describes an eccentric orbit.  These impacts occur during periapsis approach and retreat,
and the intervals between the flares are modulated by the precession of the orbit and the
emission of gravitational waves~\cite{valtonen2008massive,dey2018authenticating,laine2020spitzer}.

In this scenario, the secondary BH punches
through the geometrically thin, optically thick and radiation-dominated accretion disk of the primary BH with hypersonic velocity, shocking gas and generating hot bubbles of plasma on each side of the 
disk that expand, cool down adiabatically and eventually 
radiate by thermal bremsstrahlung after becoming 
optically thin~\cite{lehto1996oj,sundelius1997numerical},
producing the observed flares~\cite{1996ApJ...460..207L}. The time delay between the emergence of a 
bubble at the impact site and the epoch when it becomes 
transparent is an important aspect of the
binary BH model. While this can be 
estimated via detailed astrophysical modelling of the disk 
impact shock and outflow evolution \cite{pihajoki2016black}, 
it necessarily introduces some uncertainty into the 
calculations.

The
observations constrain the primary BH to have a mass $\sim 18.35 \times 10^9$ $M_\odot$ \cite{dey2018authenticating}, 
with the secondary BH having a mass
$\sim 150 \times 10^6$ $M_\odot$. The orbit of the secondary BH has an eccentricity of $\sim 0.65$, with a periapsis
$\sim 9$ and an apoapsis $\sim 48$ times the primary's Schwarzschild radius, which is $\sim 360$~AU. These properties make OJ~287 a very powerful laboratory for
probing general relativity (GR) and other aspects of fundamental physics.
The GR predictions for the binary system have been calculated to 4.5th post-Newtonian order, including the dissipative effects of the
emission of gravitational waves, and these calculations have been used to predict
successfully the time of the 2019 burst of electromagnetic
emission, which arrived within a few hours of the predicted time ~\cite{dey2018authenticating}.



In this paper we use the successful comparison of GR predictions with
the data on this unique astrophysical system to
provide for the first time a constraint on physics beyond the Standard Model of particle physics, specifically on models of dark matter. It was argued in \cite{gondolo1999dark} that if cold dark matter is present at the centre of a galaxy, 
a massive central BH distributes it adiabatically into a `spike' with a radial profile of the form
\begin{equation}
    \rho(r) \; \propto \; F (r) \left(\frac{R_{\rm sp}}{r}\right)^{\gamma_{\rm sp}} \, ,
    \label{spike}
\end{equation}
with $F (r) = (1 - 2R_S/r)^3$, where $R_S$ is the Schwarzschild radius of the BH, and $R_{\rm sp}$ scales the size of the spike. (See~\cite{2001PhRvD..64d3504U, 2002PhRvL..88s1301M} for a critique of the possible existence of a spike.)
The density slope parameter, $\gamma_{\rm sp}$, is expected to lie between 2.25 and 2.5.   The total mass of cold dark matter in the spike is obtained by integrating
(\ref{spike}) over $r$ until the density sinks below the normal galactic dark matter density, and is unknown {\it a priori} because of the unknown normalisation
of $F(r)$ that depends on the BH being considered
and its environment.  

Evidence for the 
existence of such a spike has been sought in the centre of the Milky Way. In particular, measurements of the orbit of the star S2, whose orbit samples
the gravitational field of Sgr~A* down to radii $\sim 10^{-2}$~pc,
were found in~\cite{lacroix2018dynamical} to constrain the spike mass to $\sim 1$~\% of the mass
of Sgr~A*, assuming $\gamma_{\rm sp} = 7/3$ and $R_{\rm sp} = 100$~pc,
the upper limit obtained from VLT measurements of the orbit of S2.
(A weaker constraint is provided by the concordance between the estimates of the mass of Sgr~A* based on these orbital measurements and the radius of the innermost stable circular orbit measured
by the Event Horizon Telescope, $r_{\text{ISCO}} \simeq
50$~$\mu$as~\cite{EventHorizonTelescope:2022xnr}, 
which requires that the spike mass be
$\lesssim 10$~\% of the mass of Sgr~A*.)

The eccentricity of the orbit of the secondary BH
in the OJ~287 system implies that it is sensitive to the density of any such spike
around the primary BH at radii between the binary separations at periaspsis 
and apoapsis, potentially causing deviations from the GR predictions.
We use simulations to model these deviations and fit the OJ~287 data with
modified parameters that include dark matter spikes of different magnitudes. 
We treat the gravitational effects
of the spike in the Newtonian approximation,
assuming that it is centred on the primary black hole and spherical, that shells with radii less than
the separation between the primary and secondary
have the same effect as a pointlike mass co-located
with the primary, and that the effects of shells
with radii larger than the separation can be neglected. We follow~\cite{lacroix2018dynamical} in assuming $\gamma_{\rm sp} = 7/3$, while we fix $R_{\rm sp}$  at $5 R_S$ (of the primary BH) throughout the analysis, and allow $\rho_{\rm sp}$ to vary such that the total DM mass enclosed within a region $2R_{S}\lesssim r \lesssim 50 R_{S}$ (encompassing the trajectory traversed by the secondary BH around the primary from periapsis to apoapsis) equals the studied $m_{\rm sp}/m_{1}$ ratio (ranging from 0 to 0.03). This is well within the gravitational radius of influence of the primary BH (defined as the radius at which the enclosed DM mass is twice the mass of the primary BH).
We neglect dynamical friction and post-Newtonian effects due to the spike,
which is justified {\it a posteriori} by the upper limit on its mass
that we establish.




We use the Post-Newtonian (PN) scheme to calculate the orbital dynamics up to 4.5PN order.  We work in the centre-of-mass (CoM) frame, 
converting the two-body problem to an effective one-body problem, and use harmonic coordinates. The harmonic condition, 
(de Donder gauge) is imposed, breaking general covariance.
The relative two-body acceleration and the precessional dynamics of the spin of the primary, written in terms of the unit vector $\mathbf{s}_1$, can be estimated using

\begin{eqnarray}
\label{eq:schematic_EOM}
\ddot{\mathbf{x}} \equiv \frac{d^2 \mathbf{x}}{dt^2}&=& \ddot{\mathbf{x}}_{0} + \ddot{\mathbf{x}}_{1\mathrm{PN}} + \ddot{\mathbf{x}}_{2\mathrm{PN}}  + \ddot{\mathbf{x}}_{3\mathrm{PN}}\nonumber   \\
\vspace{-0.5cm}
&+& \ddot{\mathbf{x}}_{2.5\mathrm{PN}} + \ddot{\mathbf{x}}_{3.5\mathrm{PN}} + \ddot{\mathbf{x}}_{4\mathrm{PN}\mathrm{(tail)}} +\ddot{\mathbf{x}}_{4.5\mathrm{PN}}  \\
&+&\ddot{\mathbf{x}}_{\mathrm{SO}} +\ddot{\mathbf{x}}_{\mathrm{SS}} +\ddot{\mathbf{x}}_{\mathrm{Q}} +\ddot{\mathbf{x}}_{\mathrm{4PN(SO-RR)}}\nonumber \\ 
\frac{d\mathbf{s}_1}{dt}&=& (\mathbf{\Omega}_{\mathrm{SO}} + \mathbf{\Omega}_{\mathrm{SS}} + \mathbf{\Omega}_{\mathrm{Q}}) \times\mathbf{s}_1\nonumber,
\end{eqnarray}
where $\mathrm{\mathbf{x}} = \mathrm{\mathbf{x_1}} - \mathrm{\mathbf{x_2}}$ is the CoM relative separation vector between the BHs with masses $m_{1}$
and $m_{2}$, $\ddot{\mathbf{x}}_{0}=-\frac{Gm}{r^3} \mathrm{\mathbf{x}}$ is the zeroth-order Newtonian acceleration, where $m$ is the total mass of the binary and $r = |\mathrm{\mathbf{x}}|$,
and $\mathbf{S}_1 = \frac{G m_{1}^{2}\chi_{1}}{c} \mathbf{s}_1$ is the spin of the primary with the Kerr parameter $\chi$ $\in[0,1]$ in GR. The terms $\ddot{\mathbf{x}}_{\mathrm{SO}}$ and $\ddot{\mathbf{x}}_{\mathrm{SS}}$ denote the spin-orbit (SO) and spin-spin (SS) couplings in GR, entering at 1.5PN and 2PN at leading orders, while $\ddot{\mathbf{x}}_{\mathrm{Q}}$ is the classical spin-orbit coupling (Q) arising from the quadrupole deformation of a Kerr BH, at 2PN order, and the $\mathbf{\Omega}_i$ are the SO, SS and Q contributions to the precession
of the primary BH spin vector.
Our calculations of the various terms in~(\ref{eq:schematic_EOM}) are based on an extensive body of work~ \cite{iyer1993post,iyer1995post,jaranowski1997radiative,konigsdorffer2003binary,itoh2009third, gopakumar1997second} and
will be set out in detail in an upcoming publication.




The baseline binary BH model has 9 relevant parameters:
the two BH masses $m_{1,2}$, the primary BH Kerr parameter $\chi_1$, 
the initial apocentre eccentricity $e_0$, 
the initial semimajor axis $a$ and its angle of orientation $\theta_0$, 
an ambiguity parameter $\gamma$ for the leading-order hereditary contributions to GW
emission in the BBH dynamics implemented following Equation (5) in \cite{dey2018authenticating}, an azimuthal angle $\theta_{S1}$ and a
polar angle $\psi_{S1}$ parametrising the
orientation of the primary spin vector. 

\if
The relevant parameters are determined by first computing
an approximate initial orbit via the numerical integration of the 4.5PN-order coupled ordinary differential equations 
using trial values for the independent parameters. The integration is 
performed using the LSODA algorithm \cite{petzold1983automatic}, implemented in {\tt scipy.integrate.odeint} \cite{ahnert2011odeint}.
We use the resultant orbit to extract a list of reference times at which the 
secondary BH crosses the accretion disk of the primary, which in practise we 
assume to lie in the $z = 0$ plane. 
\fi

When making fits, the plane-crossing epochs must be corrected to
incorporate astrophysical processes between 
an impact and an observed optical outburst, with a time delay $t_{\mathrm{del}}$ added to 
account for the time lag between the generation of the plasma bubbles and the 
epoch at which they become optically thin. An additional timing correction
$t_{\mathrm{adv}}$ is applied to model the tidal force of the approaching secondary 
BH that warps the disk and advances the impacts.
We adopt empirical models of the time delay and advance based on the
best-fit orbit of \cite{dey2018authenticating},
which our good fits resemble. In our orbital fitting procedure, the observational uncertainties are
all assumed to be Gaussian and uncorrelated, and
the optimisation algorithm adopted is the Nelder-Mead algorithm \cite{nelder1965simplex}, which
we implement using the \href{https://github.com/alexblaessle/constrNMPy}{\texttt{constrNMPy}} package.



\begin{figure}[htb]
    \centering
    \vspace{-5mm}
    \includegraphics[width=0.5\textwidth]{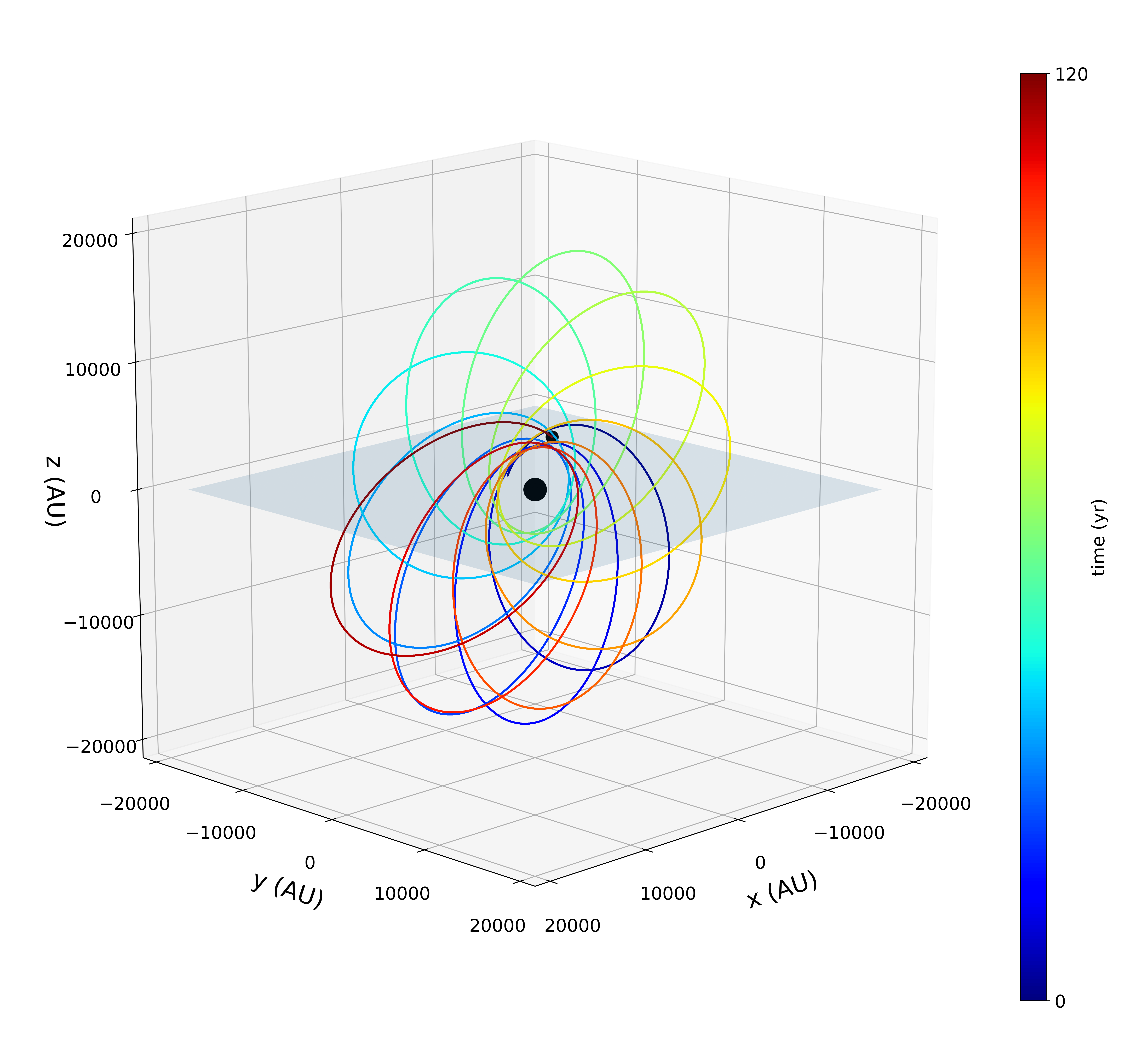}
    \caption{\it
    The precessing orbit of the secondary BH (smaller black spot) in a coordinate system
    centred on the primary BH (larger black spot).
    Note the passages of the secondary BH through the accretion disc (represented as a shaded plane).}
    \label{fig:wool}
\end{figure}

Fig.~\ref{fig:wool} displays the evolution of the OJ 287 system in our best no-DM fit over a period of 120~y,
corresponding to 10 orbits of the secondary BH (represented by the smaller black spot), in a coordinate system
centred on the primary BH (represented by the larger black spot). We see clearly the large orbital precession
and the passages of the secondary BH through the accretion disc (represented as a shaded plane)
as it approaches and retreats from the periapsis.

Table~\ref{tab:Outburst_times} lists in the first column the starting times of flares of OJ~287
starting with that at the end of 1912, with the corresponding uncertainties shown in the second column.
The timings of the flares not written in bold were not measured accurately
and are not used in our fits. Ref.~\cite{dey2018authenticating} used 10 flares, 
corresponding to 9 time differences that they fitted using the 9 parameters
listed above. In our analysis we have one additional parameter characterizing the spike, and our fits
include the well-measured 2019 flare as well as the less accurately
measured flares in 1959 and 1994.  The contribution of the spike to the dynamics is included at zeroth order and follows the mass profile obtained from (\ref{spike}).
The third column in the Table lists the predictions for the starting times in our best-fit no-DM spike model,
the fourth column is our best fit for a spike mass ratio $m_{\rm sp}/m_1 = 0.02$, and the
fifth column is that for a spike mass ratio of 0.03.

\begingroup 
\squeezetable 
\begin{table}[ht]
\centering
\textit{Flare times, uncertainties and model estimates} \\
\begin{tabular}{|c|c|c|c|c|}
\hline 
Julian year                     & uncertainty                    & No spike & 2\% spike                       & 3\% spike                       \\ \hline \hline
{\bf 1912.980}                        & ±0.020                         &  {1912.982}      & {1912.968 } & {1912.958} \\
1922.529                       & -                             &  {1922.536}      & {1922.537 } & {1922.537} \\
1923.725                       & -                             &  {1923.730}      & {1923.725 } & {1923.722} \\
1934.335                       & -                             &  {1934.337}      & {1934.340} & {1934.340} \\
1935.398                       & -                             &  {1935.403}      & {1935.402 } & {1935.400 } \\
1945.818                       & -                             &  {1945.819}      & {1945.822 } & {1945.822 } \\
{\bf 1947.283}                        & ±0.002                         &  {1947.285}     & {1947.284 } & {1947.283 } \\
{\bf 1957.095}                        & ±0.025                         &  {1957.087}      & {1957.078} & {1957.073 } \\
{\bf 1959.25}  & { ±0.05}   &  {1959.216}      & { 1959.213} & {1959.213 } \\
1964.231                       & -                             &  {1964.242}      & {1964.216} & {1964.204 } \\
1971.126                       & -                             &  {1971.129}      & {1971.127} & {1971.127 } \\
{\bf 1972.935}                        & ±0.012                         &  {1972.934}      & {1972.921} & {1972.916} \\
{\bf 1982.964}                        & ±0.0005                        &  {1982.964}      & {1982.965} & {1982.965 } \\
{\bf 1984.125}                        & ±0.01                          &  {1984.120}      & {1984.116 } & {1984.114} \\
{\bf 1994.77}  & { ±0.1}    &  {1994.595}      & {1994.599 } & {1994.599} \\
{\bf 1995.841}                        & ±0.002                         &  {1995.839}      & {1995.838 } & {1995.838} \\
{\bf 2005.745}                        & ±0.015                         &  {2005.747}      & {2005.754 } & {2005.754} \\
{\bf 2007.6915}                       & ±0.0015                        &  {2007.693}      & {2007.691 } & {2007.692} \\
{\bf 2015.875}                        & ±0.025                         &  {2015.882}      & {2015.882} & {2015.872} \\
{\bf 2019.569} & { ±0.0005} &  {2019.569}      & {2019.568} & {2019.569} \\ 
\hline 
\end{tabular}
\caption{\it \label{tab:Outburst_times} Starting epochs (in Julian
years) of the observed optical flares of OJ 287 
from 1912 onwards.
First column: The data points prior to 1970 were obtained from
archival photographic plates while the historical
1912/3 flare time is from \cite{hudec2013historical}. We only use the flares listed in bold, the starting times of the other flares are not known accurately and are not used in the fits. Second column: The uncertainties in the starting times.
Third, fourth and fifth columns: Our predictions for the starting times 
in the best-fit no-DM spike model and in the best-fit models with spike mass ratios of 2 and 3\%.}
\end{table}
\endgroup

\begin{figure}[htb]
    \centering
     \includegraphics[width=0.46\textwidth]{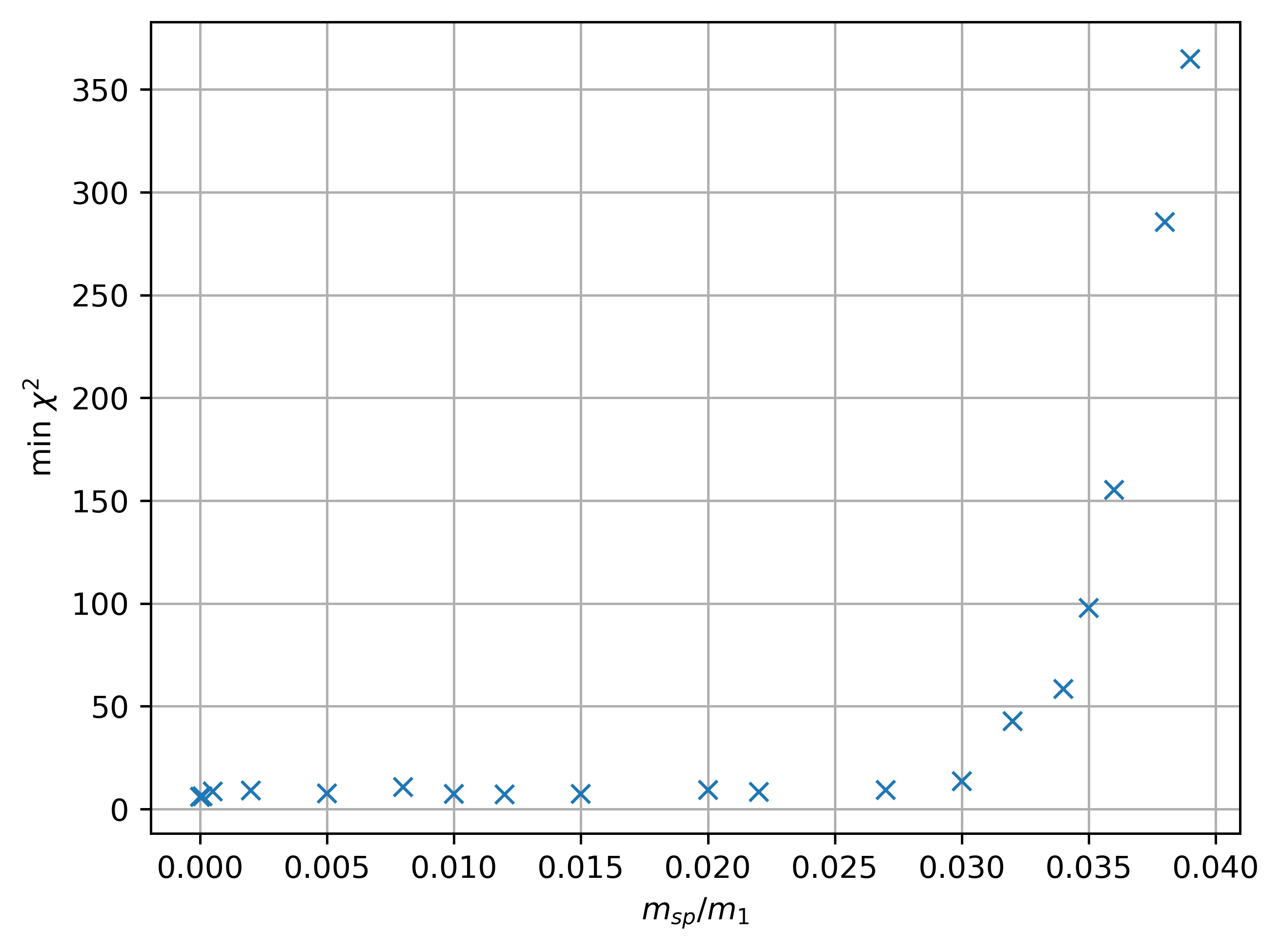}
    \caption{\it
    The dependence of the $\chi^2$ likelihood function
    on the ratio $m_{\rm sp}/m_1$, as obtained from the sampling of model
    parameters described in the text.}
    \label{fig:results}
\end{figure}

Fig.~\ref{fig:results} shows the minimum values of $\chi^2$
that we found for the
indicated choices of the spike mass ratio, $m_{\rm sp}/m_1$. The best-fit no-DM model has
$\chi^2 = 6.05$, and the best fits for values of $m_{\rm sp}/m_1 < 0.03$ all have
$\chi^2 \lesssim 10$, corresponding to $p$-values $\gtrsim 20$\%. However, the best fit for 
$m_{\rm sp}/m_1 = 0.03$ has $\chi^2 \simeq 13.6$, corresponding to a $p$-value $\simeq 3$\%,
and the values of $\chi^2$ rise sharply as $m_{\rm sp}/m_1$ increases beyond $0.03$.
For example, $\chi^2 \simeq 99$ already for $m_{\rm sp}/m_1 = 0.035$, corresponding to
a $p$-value $\simeq 10^{-18}$. On the basis of these results we conclude that the
current OJ 287 data set an upper limit $m_{\rm sp}/m_1 \lesssim 0.03$.

\begin{figure}[htb!]
    \centering
    \includegraphics[width=0.418\textwidth]{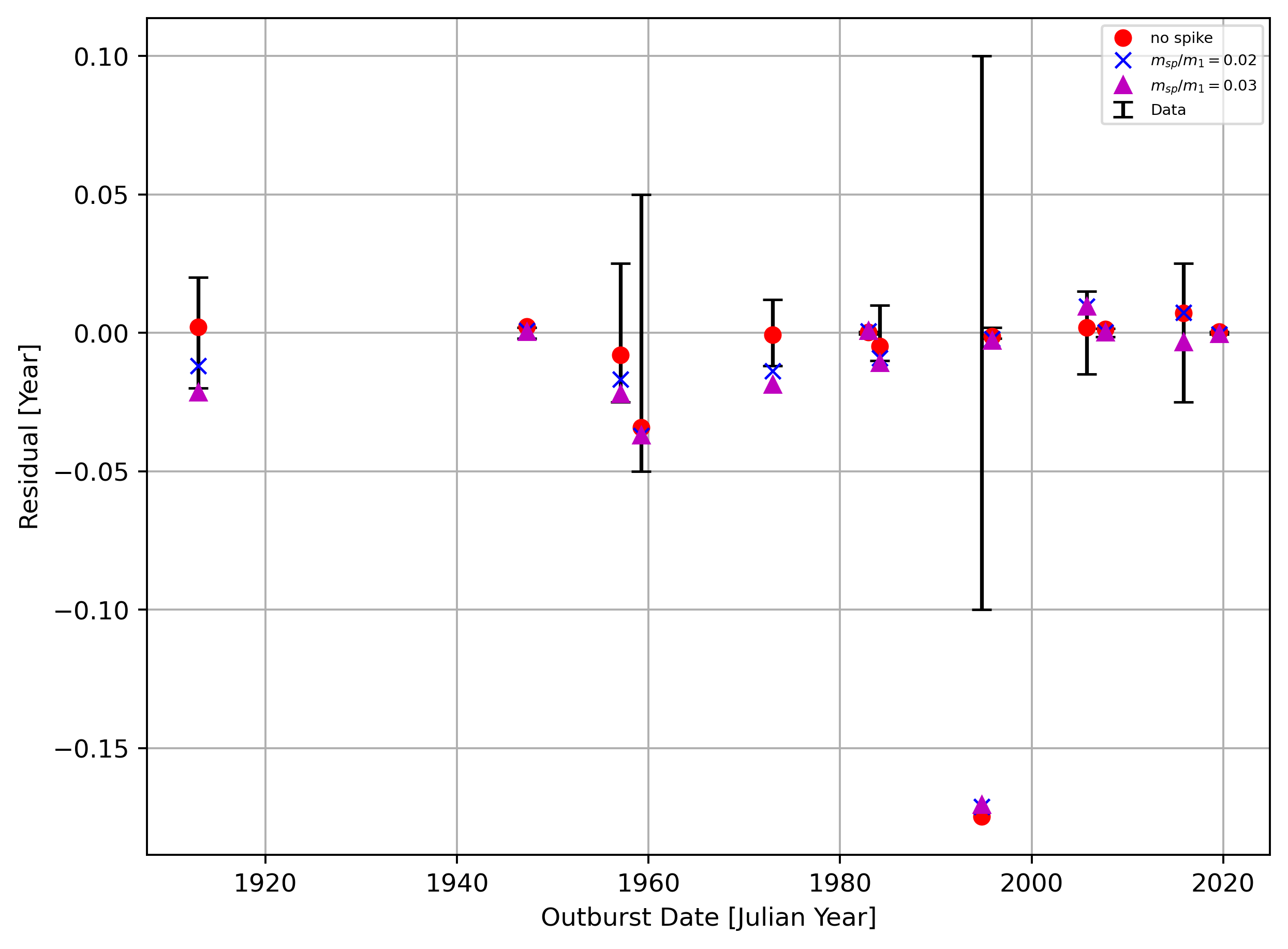}
      \includegraphics[width=0.432\textwidth]{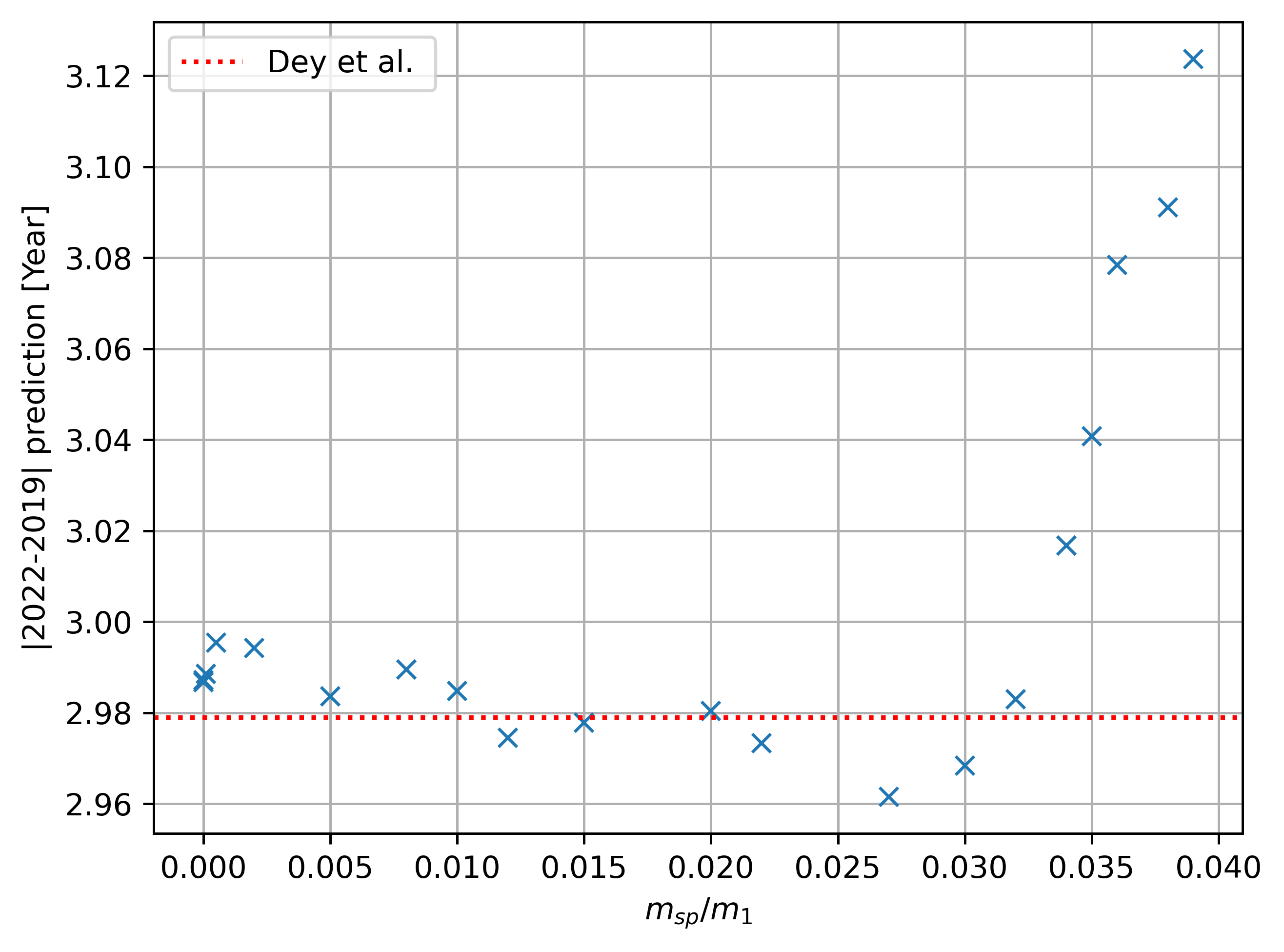}
    \caption{\it
    Upper panel: The best-fit predictions for the flare timings calculated in
    a model without a dark matter spike (red spots), as obtained from the sampling of model
    parameters described in the text, compared with the measured timings (black).
    Also shown are the timings for the best fits with $m_{\rm sp}/m_1 =0.02$ (blue crosses)
    and $m_{\rm sp}/m_1 =0.03$ (purple triangles).
    Lower panel: A scatter plot of the predictions for the interval
    between the 2022 and 2019 flares (vertical axis)
    for our best fits with the indicated values of $m_{\rm sp}/m_1$ 
    (horizontal axis). We also show the prediction of~\cite{Valtonen:2022lfc} for the
    the interval between the 2022 and 2019 flares (horizontal red dashed line).}
    \label{fig:predictions}
\end{figure}

The upper panel of Fig.~\ref{fig:predictions} compares the predictions of our
best-fit models with no DM spike (shown as red spots), $m_{\rm sp}/m_1 = 0.02$ 
(shown as blue crosses) and $m_{\rm sp}/m_1 =0.03$ (shown as purple triangles) with the observations
(shown in black). We see a high level of consistency for the no-DM model, the most noticeable
deviation being that for the 1994 flare. However, the observations of this flare
were the least precise among the flares fitted, and its timing is the most
uncertain, so we do not regard this deviation as being significant. There is also
a good level of consistency for the best fit with $m_{\rm sp}/m_1 =0.02$.
However, the best fit with $m_{\rm sp}/m_1 =0.03$ clearly does not fit the data as well,
enabling the eye to confirm the numerical analysis presented in the previous paragraph.

Our no-DM spike prediction for the
very well measured 2019 flare shown in Table~\ref{tab:Outburst_times} agrees with the measured timing 
to within a few hours, within the uncertainty
in the starting time of the 2019 flare. The timing of the 1982 flare is
as well known as that in 2019, and our no-DM spike prediction for it again 
agrees with the measurement within a few hours.

As further support for our conclusion that
$m_{\rm sp}/m_1 \lesssim 0.03$, we have considered two possible side-effects of interactions between the
secondary black hole and the spike. One is the dissipative effect of back-reaction on the orbit of the secondary black hole, and the other is the possible disruption of the spike by the secondary black hole. Following the approach of~\cite{Kavanagh:2020cfn}, we find that both effects can be neglected in a first approximation.

According to the model of~\cite{valtonen2008massive,dey2018authenticating,laine2020spitzer}
another flare of OJ~287 is expected shortly, though there are considerable
astrophysical uncertainties in its timing, as discussed
in~\cite{Valtonen:2022lfc}.
We now discuss how the timing of this flare
might differ from the no-spike pure GR case
discussed in~\cite{Valtonen:2022lfc}. As seen in the lower panel
of Fig.~\ref{fig:predictions}, our estimates for 
$m_{sp}/m_1 \lesssim 0.03$ suggest that the next flare might be advanced 
by $\lesssim 5$~days relative to the no-spike prediction. On the other hand, 
we estimate that it would be delayed by $> 10$~days for $m_{sp}/m_1 \ge 0.035$. 
In view of the uncertainty in the arrival time estimated in~\cite{Valtonen:2022lfc}, 
our results suggest that observations of the next flare of OJ~287 may not be able to strengthen our
constraint $m_{sp}/m_1 \le 0.03$ based on the timings of previous flares.

GR has already been tested by observations of OJ~287 to the 4.5PN order, including the effects of
GW emission \cite{Spitzer}, and will be tested again by observations of future flares. We have shown in the paper, for the first
time, how
observations of OJ~287 can be used to constrain models of new physics
beyond the Standard Model of particle physics. Specifically,  observations of
OJ~287 can constrain the possible existence of a cold dark matter
spike surrounding its primary BH. They already constrain the possible mass
of such a spike to $m_{\rm sp}/m_1 < 0.03$ at the 99.9\% CL, and measurements of the timing of the expected future flares
and more detailed modelling have the potential to constrain further the spike mass.

There are certainly ways in which the modelling of dark matter effects on the evolution
of OJ~287 could be improved. For example, we have considered the possible static gravitational effects of a cold dark matter
spike but have not considered other possible interactions between the secondary black hole and the spike,
such as tidal forces or non-gravitational interactions. Another important area for future research
will be to refine the modelling of the dynamical interactions between the secondary black hole
and the accretion disk, and the time delay between the nominal impact on the accretion disk and
electromagnetic emissions from the cooling of the plasma bubble that it generates,
which is the dominant uncertainty in the prediction of the timing
of any future flare.

OJ 287 has barely begun to demonstrate its potential for probing not just astrophysics and GR, but also other aspects of fundamental physics.

\section*{Acknowledgements}

We are grateful to Mauri Valtonen and Pauli Pihajoki for fruitful discussions and valuable comments.  
The work of AA was supported by a UK STFC studentship,
that of JE and MF was supported in part by STFC Grants ST/P000258/1 and ST/T000759/1, and that of JE also
in part by the Estonian Research Council via a Mobilitas Pluss grant.

\label{Bibliography}
\bibliographystyle{unsrtnat}  
\bibliography{Bibliography}  

\end{document}